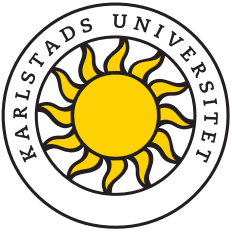

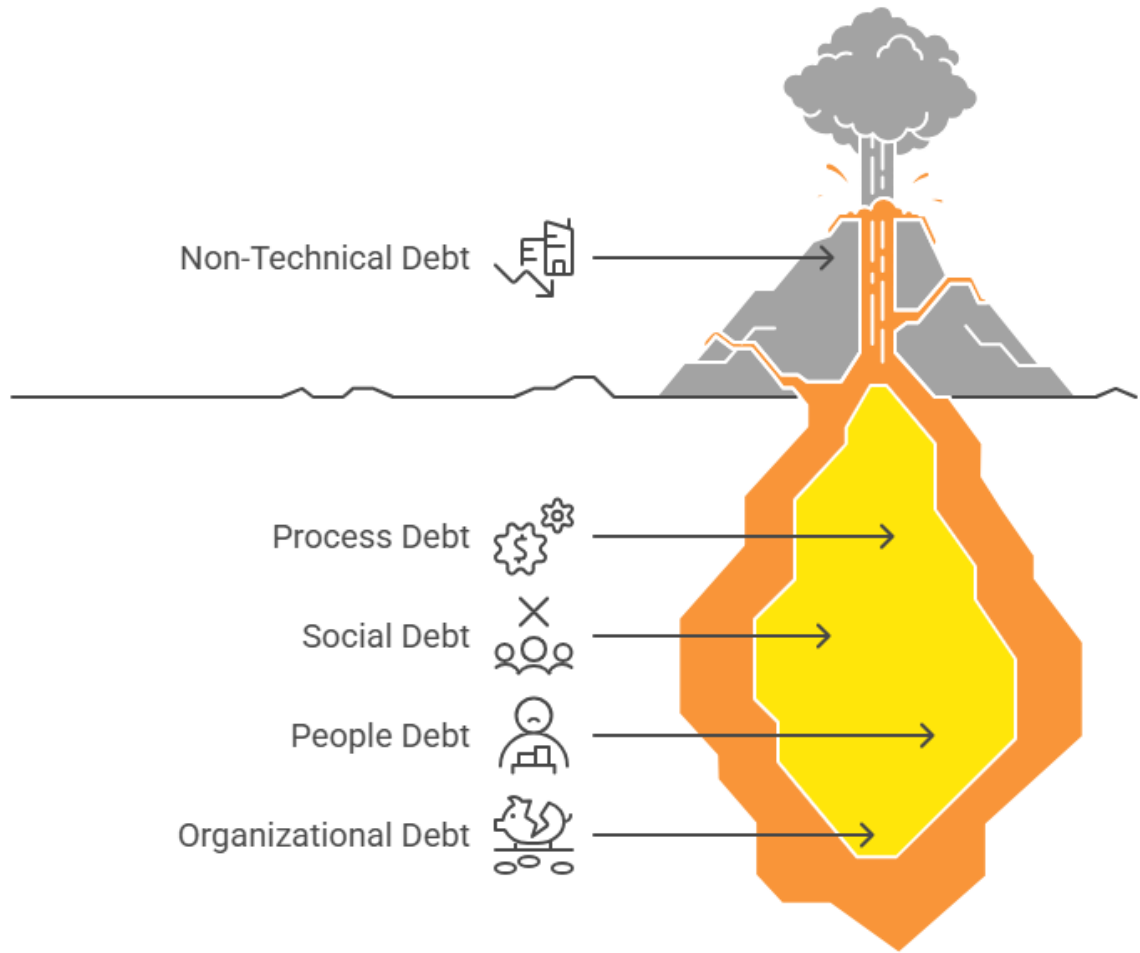

# NON-TECHNICAL DEBT IN AGILE SOFTWARE DEVELOPMENT

Insights from the NODLA Research Project (2021–2024)

---

Muhammed Ovais Ahmad & Tomas Gustavsson
Karlstad University, Sweden

KAU.SE

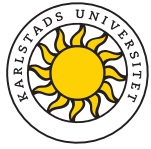

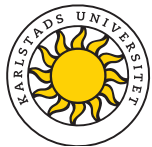

**NON-TECHNICAL DEBT IN AGILE SOFTWARE DEVELOPMENT**
INSIGHTS FROM THE NODLA RESEARCH PROJECT (2021–2024)

Muhammed Ovais Ahmad & Tomas Gustavsson

*Karlstad University, Sweden*

2025

# EXECUTIVE SUMMARY

Non-Technical Debt (NTD) is a common challenge in agile software development, manifesting in four critical forms:

- *Process* Debt arises from inefficient or outdated workflows that hinder agility and adaptability. Examples include misaligned processes, poor synchronization across teams, and unclear role definitions, all of which can slow progress.
- *Social* Debt stems from suboptimal team dynamics or organizational culture, such as poor communication, lack of trust, or fixed silos. These issues hinder collaboration, increase misunderstandings, and often result in costly rework.
- *People* Debt refers to issues with people and their competence. It reflects challenges related to human resources and expertise, such as inadequate training, hiring delays, or overworked teams. This form of debt limits an organization's ability to retain skilled, motivated personnel and meet increasing demands.
- *Organizational* debt arises from outdated structures, policies, or practices that no longer align with the organization's goals. Such rigidity limits innovation, hinders adaptability, and prevents the pursuit of operational excellence.

The NODLA project1[1] (2021–2024), a collaboration between Karlstad University and four leading Swedish industrial partners, reveals how various debt types disrupt large-scale Agile Software Development (ASD) environments. Through extensive surveys, in-depth interviews, and statistical analyses involving a diverse group of software professionals, we identified key drivers of NTD and their impacts. Our findings emphasize:

- Well-structured, highly cohesive teams learn faster, adapt more effectively, and innovate consistently.
- Psychological safety, fostered by proactive leadership, is essential for innovation, experimentation, and keeping employees.
- Inefficient processes and unclear roles contribute significantly to drops in job satisfaction, productivity, and team morale.
- Social fragmentation, particularly in remote and hybrid settings, breeds rework, delays, and increased costs.
- Neglected human resource needs, such as delayed hiring or insufficient training, limit an organization's ability to meet growing demands.

This white paper distils these insights into practical, evidence-based strategies, such as refining team composition, clarifying roles, fostering psychological safety, streamlining workflows, and embracing failure as a learning tool. By implementing these strategies, organizations can reduce NTD, reclaim agility, and unlock their teams' full potential.

---

1 https://www.kau.se/cs/forskning/pagaende-projekt/nationella-projekt/nodla

# TABLE OF CONTENTS

# INTRODUCTION

In 1992, Ward Cunningham introduced the concept of "technical debt" to describe the trade-offs developers make to accelerate software delivery, increasing future costs in maintenance, rework, and performance. Over the decades, technical debt has become a cornerstone of software engineering discourse, with tools and methodologies developed to detect, measure, and mitigate it.

As agile methodologies scaled, however, a parallel challenge emerged: Non-Technical Debt (NTD). Unlike technical debt, which is often visible in code (e.g., poor documentation, hardcoded values), NTD is treacherous, lurking in team dynamics, workflows, structures, and human practices. It encompasses the social, process, people, and organizational factors that, when neglected, reduce the effectiveness of agile teams. Left unchecked, NTD can damage collaboration, limit innovation, increase costs, and undermine long-term success.

**The NODLA Project (2021–2024)**

The NODLA initiative was a collaboration between Karlstad University and four leading Swedish industrial partners, launched to tackle NTDs head-on. Over the course of three years, we examined NTD in the context of large-scale Agile Software Development (ASD) environments, blending academic rigor with industry pragmatism. Our research involved:

- **Interviews:** In-depth discussions with diverse groups of software professionals and leaders to uncover qualitative insights into NTD manifestations and root causes.
- **Surveys:** Diverse groups of software professionals across multiple organizations, using validated instruments to measure NTD impacts on job satisfaction, team performance, and innovation.
- **Statistical Analyses:** Quantitative studies to identify correlations and causal relationships, such as the link between team structure and knowledge sharing or psychological safety, performance, and learning.

Our goal was to identify NTD root causes, measure their impacts, and develop actionable, evidence-based strategies to mitigate them. This white paper synthesizes our findings, providing a comprehensive examination of NTD manifestations and potential solutions.

# ORGANIZATIONAL DEBT INSIGHTS

Organizational Debt can paralyze even the most talented teams, stifling learning, adaptability, and innovation. Organizational Debt creates systemic barriers to excellence, often reinforcing other forms of NTD. These forms of NTD are interconnected, often reinforcing one another. For instance:

- Unclear roles (Process Debt) can strain team relationships (Social Debt), as members struggle to align their efforts.
- Overburdened staff (People Debt) may lack the bandwidth to adapt to rigid structures (Organizational Debt), leading to burnout.
- Poor communication (Social Debt) can exacerbate process inefficiencies (Process Debt), creating a vicious cycle of delays and rework.

Addressing NTD requires a holistic approach, recognizing their interplay and cumulative impacts. Our studies reveal two critical levers for countering Organizational Debt in a sustainable way: knowledge sharing and psychological safety.

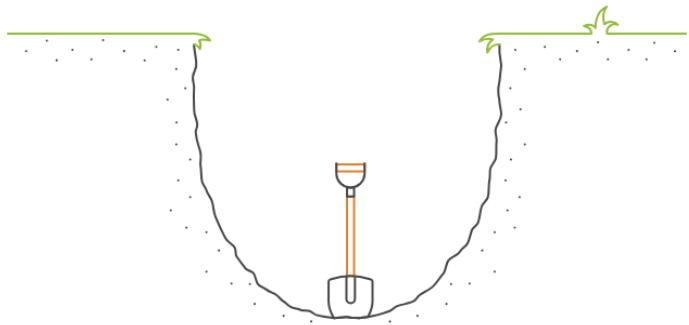

Organizational Debt stifles learning and innovation.

### Knowledge sharing – The engine of learning and adaptability

In a quantitative study involving 159 practitioners across five Swedish companies, we explored how team-level factors influence learning in agile environments [8]. Using statistical analyses, we identified team structure as a key driver of cohesion and reflexivity, both essential for effective knowledge sharing. Key findings:

- Cohesive teams, where skills and roles were balanced, demonstrated faster problem-solving and reduced redundant work.
- Successful learning in large-scale ASD projects requires a suitable learning environment that empowers software professionals to interact frequently and openly, reflect on their actions, and share their knowledge.
- Learning environment is shaped by team structure and team cohesion. The learning environment, in turn, impacts knowledge sharing and team reflexivity (i.e., learning activities), which contribute to learning in large-scale ASD projects.

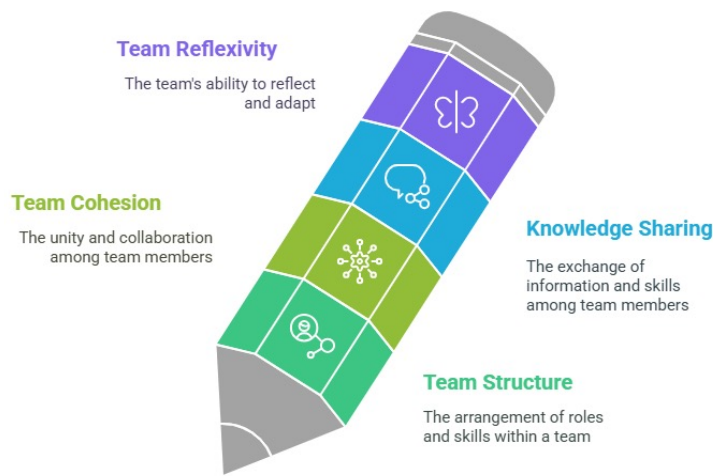

Figure 1. Studied factors and their impact on team learning.

Knowledge sharing is the lifeblood of agile teams. It accelerates problem-solving, reduces redundant work, and fosters innovation. However, without a supportive structure, teams struggle to share insights effectively, leading to siloed knowledge, missed opportunities, and inefficiencies. In large-scale ASD, where coordination is complex, poor knowledge sharing can derail entire projects.

**Actionable strategies**
- **Team composition is crucial; dare to question it.** Pay attention to team composition and consider skill sets, personality types, and how individuals can complement each other. Although Agile methods emphasize that teams should not be changed, our study reveals that without a well-structured team, building trust and collaboration among team members with low team cohesion is challenging.
- **Strong team cohesion drives reflection and adaptation.** Strong team cohesion should be prioritized. It enhances the team's ability to reflect on their work and improves the chances that the team actually questions and adapts their processes.
- **Knowledge sharing is critical for learning.** Encourage team members to actively share knowledge and experiences and implement processes for continuous knowledge sharing. Knowledge sharing is even more crucial than team cohesion when it comes to improving the team's learning ability. Teams that prioritize knowledge exchange are more likely to solve problems faster and avoid redundant work.
- **Leverage technology.** Use tools like Confluence, Slack, or Microsoft Teams to document and disseminate knowledge, ensuring accessibility across distributed teams.

It is essential to regularly measure and adjust your organization. For example, deploy a quarterly survey with questions like, "How often do you share insights with colleagues?" to identify gaps.

## Psychological Safety – The foundation of innovation and retention

In our case study [5] with agile teams, we assessed the role of psychological safety, defined as the freedom to voice ideas without fear, in fostering innovation and retention. Proper leadership is key to foster psychological safety. Key findings:

- Teams with leaders who showed vulnerability (e.g., admitting mistakes) and encouraged feedback have positive impact on work.
- Psychological safety was strongly correlated with employee retention.
- Teams lacking psychological safety avoided experimentation, fearing blame or reprisals, which stifled creativity and adaptability.

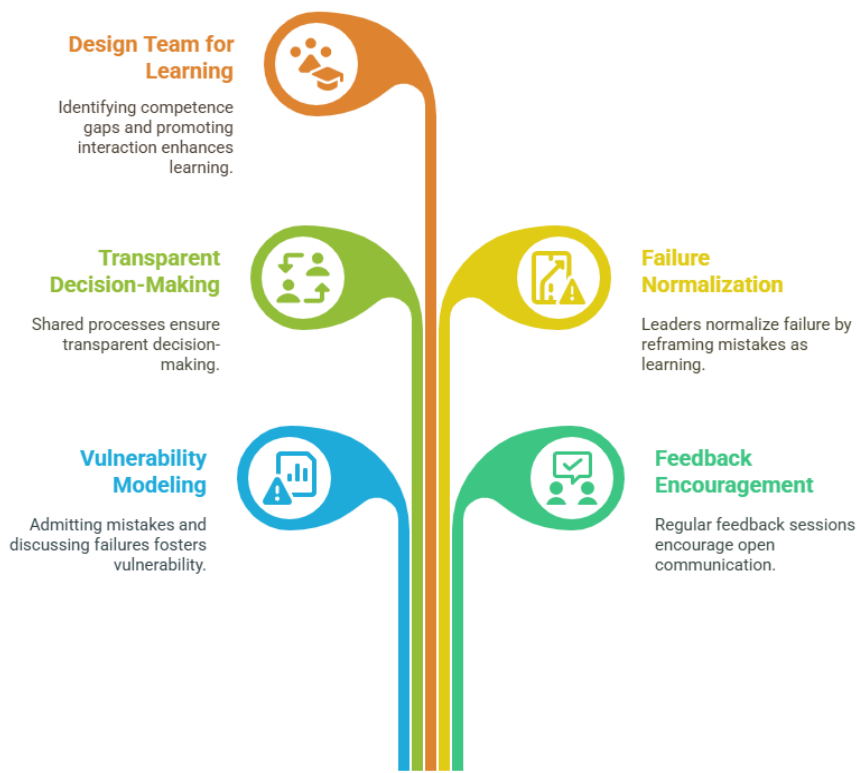

Figure 2. Identified leadership principles essential for creating a psychologically safe work environment.

Psychological safety is the foundation of learning culture. It enables teams to take risks, experiment with new approaches, and retain top talent. Without it, organizations face "brain drain," as dissatisfied employees seek more supportive environments.

### Actionable strategies

- **Foster psychological safety with proactive leadership.** Leaders must create an environment where team members feel comfortable expressing their thoughts without fear of criticism. This includes promoting openness in decision-making and feedback processes.

- **Acknowledge your failures to build psychological safety.** By presenting your own failures transparently, you create a culture of vulnerability where team members feel safe to admit their own mistakes. This fosters a non-judgmental environment that encourages risk-taking, experimentation, and innovation. For instance, a Scrum Master might share, "In our last sprint, I underestimated the complexity of this feature–here is what I learned."
- **Emphasize teamwork values during the onboarding process.** Discuss teamwork, work culture, and the importance of mutual respect and trust among team members in the onboarding process. For example, include a module on "Fostering a Safe Team Environment" in the onboarding process.
- **Measure Psychological Safety.** Utilize validated instruments, such as Amy Edmondson's Psychological Safety Scale, to regularly track team health. Questions like "Do you feel comfortable raising concerns without fear?" can identify areas for improvement.

### A case example: Specialized Roles and Cultural Initiatives

A case company introduced two specialized roles in their Agile way of working: Product Guardian and Security Master. Each Product Guardian focused on code quality and architectural integrity in one specific functional area (i.e. product). Security Masters were dedicated to embedding security throughout the development lifecycle. These roles also cultivated a culture of continuous learning, team autonomy, and psychological safety.

**Product Guardians:** These experts ensured coding standards were upheld and technical debt was managed. Their responsibilities included:

- Enforcing consistent coding practices across teams.
- Conducting detailed code reviews to catch defects early.
- Providing architectural oversight to maintain system scalability.
- Facilitating knowledge sharing through cross-team workshops.

**Security Masters:** These specialists integrated security from the project's outset, with duties such as:

- Performing regular security assessments.
- Training developers in secure coding practices.
- Collaborating with Product Guardians to align security with architectural goals.
- Ensuring compliance without disrupting agile workflows.

Both roles operated collaboratively within Scrum ceremonies, advising teams without imposing excessive control. For instance, during sprint planning, Product Guardians aligned features with architectural standards, while Security Masters flagged security considerations. This ensured quality and security were proactive priorities rather than reactive fixes.

The case company leadership complemented these roles with a number of cultural initiatives:

- **Continuous Learning:** Biweekly cross-team demos and a shared wiki encouraged knowledge sharing, reducing silos.
- **Team Autonomy:** Teams were empowered to make decisions and take risks, fostering innovation within clear accountability structures.
- **Psychological Safety:** Leaders promoted an environment where mistakes were learning opportunities, using "failure retrospectives" to extract insights.
- **Double-Loop Learning:** Beyond fixing immediate issues (single-loop learning), teams analysed underlying processes (double-loop learning) to prevent recurrence. For example, a security flaw prompted not just a patch but a review of coding practices.

These efforts created a resilient, adaptive workforce capable of tackling the complexity in a large-scale agile software development environment.

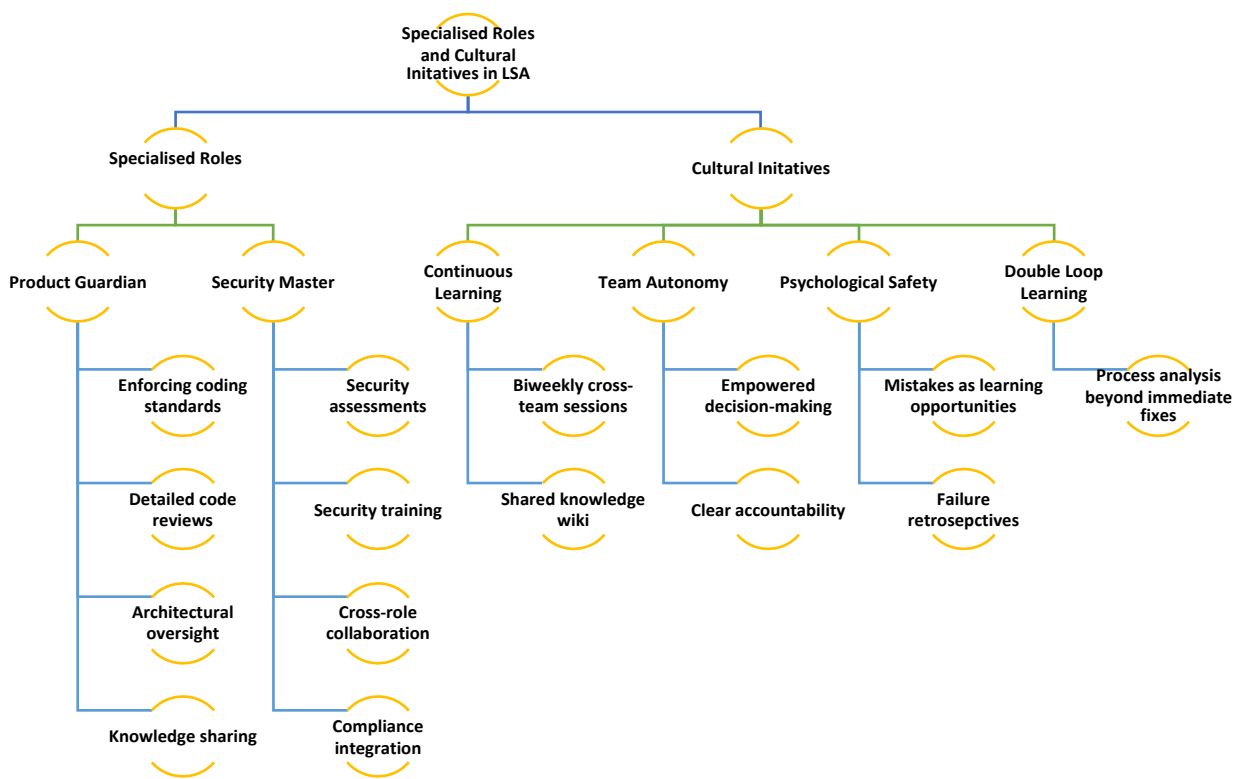

**Figure 3.** An overview of implemented changes in the case company.

# PROCESS DEBT INSIGHTS

Process Debt drags down efficiency, morale, and productivity. Our research identifies two key areas of concern: job satisfaction and process inefficiency.

## Job Satisfaction – The hidden cost of workflow inefficiencies

In our quantitative survey of 191 developers, using a validated instrument, we explored the link between Process Debt and job satisfaction [2]. Key findings:

- Process Unsuitability (e.g., outdated workflows) and Roles Debt (e.g., unclear responsibilities) accounted for 33.8% of the variance in job satisfaction.
- Other factors, such as Documentation Debt, showed weaker links, suggesting that process-related frustrations cut deeper than technical inefficiencies.
- Qualitative interviews revealed that unclear roles led to frustration, overload, and disengagement.

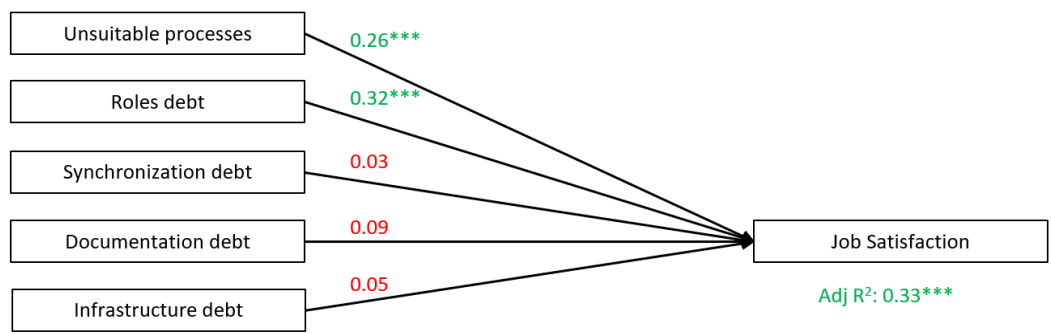

Figure 4. Studied PD types and their impact on job satisfaction. Asterisks (*) show significant relationships.

Dissatisfied teams are less productive, less engaged, and more prone to turnover. While technical inefficiencies may slow progress, process-related frustrations, such as misaligned workflows or unclear roles, more profoundly affect morale and motivation. In large-scale ASD, where teams rely on synchronized efforts, Process Debt can ruin entire projects.

**Actionable Strategies**

- **Clarify roles through workshops.** Workshops or brief alignment meetings can help ensure everyone understands their responsibilities and how they contribute to broader goals. For example, conduct a quarterly role-mapping workshop where team members define and document their responsibilities.
- **Leverage retrospectives.** Use sprint retrospectives to identify and address workflow pain points, adapting processes iteratively. For instance, dedicate 10 minutes of each retrospective to discussing, "What processes slowed us down this sprint?"
- **Streamline workflows with tools.** Implement tools like Jira or Trello to visualize workflows, identify bottlenecks, and ensure alignment across teams.
- **Monitor with pulse surveys.** Deploy short, validated questionnaires quarterly to track Process Debt [3] and guide interventions.

## Process Inefficiency – The ripple effect on technical debt and costs

In-depth interviews [4] with software professionals revealed that process inefficiencies, such as poorly managed meetings, inadequate sprint planning, or misaligned cross-team workflows, lead to delays, increased costs, and low morale. Key insights:

- Inefficient processes often forced teams to cut corners, directly contributing to TD. For example, rushed sprint planning led to skipped testing, resulting in buggy releases.
- Poorly managed meetings waste developer coding time, inflate costs, and cause frustration.
- Misaligned cross-team workflows created bottlenecks, causing teams to wait idly for dependencies to be resolved.
- The lack of clarity in roles, such as a missing mediator or overloaded Scrum Masters, further hindered team productivity and coordination.

Process Debt doesn't just slow teams, it creates a cascade of problems from duplicated efforts to suboptimal code. In large-scale ASD, where synchronization is crucial, inefficiencies can ripple across teams, increasing costs and compromising code quality. Addressing Process Debt is essential for both short-term efficiency and long-term success.

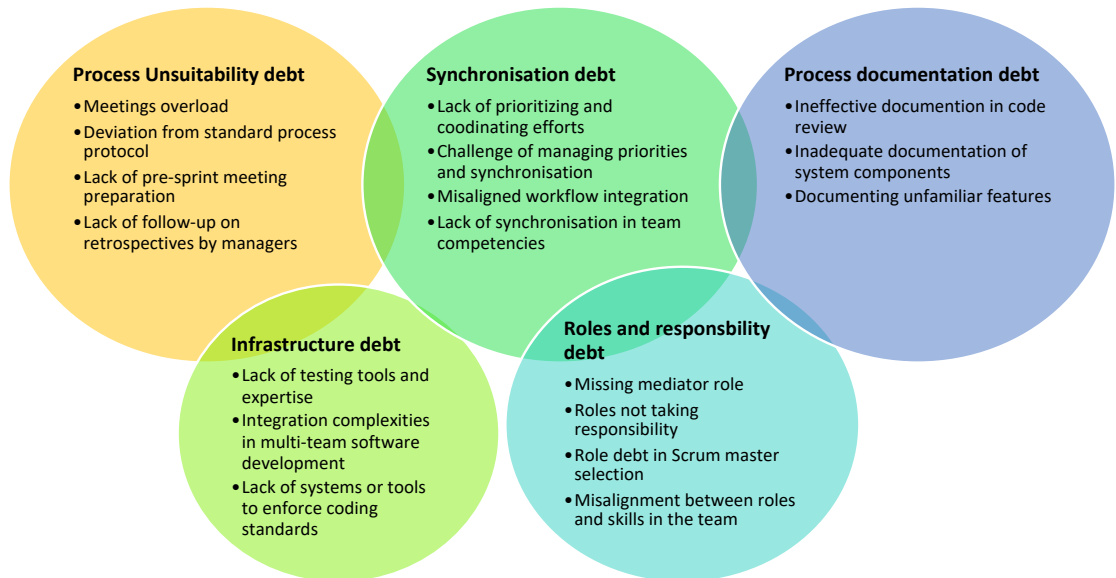

Figure 5. Identified process debt types and examples of subtypes in large-scale Agile software development.

## Actionable Strategies

- **Promote cross-functional collaboration.** Foster a collaborative culture among teams with different areas of expertise, ensuring that team members from diverse backgrounds can work together effectively. The cross-functional collaboration will close knowledge gaps, leading to more efficient work.
- **Clarify roles and responsibilities.** Provide clear definitions for all roles in the organization, particularly for Scrum Masters, Product Guardian, Security Master, Product Owners, and coordinating roles. This will prevent overload and confusion.
- **Streamline cross-team workflows.** Implement tools or rituals (e.g., shared Kanban boards, daily syncs) to synchronize efforts across silos. For example, use a shared board to visualize dependencies and track progress.
- **Optimize meetings:** Set clear agendas, time limits, and outcomes for meetings. For example, limit daily stand-ups to 15 minutes and focus on blockers and next steps.
- **Treat Process Debt as a priority:** Use retrospectives to identify and tackle process bottlenecks, just as you would Technical Debt. For instance, dedicate one retrospective per quarter to process improvement, asking, "What workflows can we streamline?"

# SOCIAL DEBT INSIGHTS

Social Debt fractures teams, undermining communication, trust, and collaboration. We identified two key areas: team dynamics and learning from failures.

## Team Dynamics – Finding the purpose

Through in-depth interviews, we identified communication breakdowns, collaboration hurdles, and coordination challenges, organizational fragmentation, and social constraints as primary drivers of Social Debt [6]. Key insights:

- Poor communication led to misunderstandings, missed requirements, and rework, which eventually inflating costs.
- Collaboration hurdles, such as organizational silos, created friction, as teams spending their time resolving coordination issues.
- Remote work intensified these issues, as some teams reporting increased isolation and misalignment in hybrid settings.

Disconnected teams face delays, rework, and frustration, all of which inflate costs and erode morale. In large-scale ASD, where coordination is complex, Social Debt can be particularly damaging, leading to missed deadlines and suboptimal outcomes.

### Actionable Strategies

- **Align the team with a shared vision.** By defining team goals and clarifying a shared vision, team members gain a deeper understanding of the purpose of their work. Changes in team composition and work assignments can gradually shift focus, and "old truths" may hinder the team from excelling in its work.
- **Create a unified vision across teams.** Implement regular alignment sessions or workshops to ensure that all teams understand and share a common vision for your work. A unified vision reduces misalignments, improves decision-making, and ensures teams work cohesively towards shared objectives, thereby reducing social debt.
- **Build remote-friendly frameworks.** Develop frameworks for effective teamwork in hybrid or fully remote settings, ensuring that all team members are integrated and valued, regardless of their location. This improves collaboration, morale, and reduces the risks of isolation that contribute to social debt. For instance, schedule weekly virtual coffee breaks to foster informal connections.
- **Measure social health.** Use team health checks or surveys to track communication quality and intervene early. Questions like, "Do you feel connected to your teammates?" can identify gaps.

## Learning from Failures – Turning setbacks into springboards for success

Our study [9] involving agile teams revealed a strong positive correlation between embracing failure and team performance. Key findings:

- Teams that openly discuss mistakes have higher adaptability.
- Psychological safety was the enabler, allowing members to experiment without fear. Teams lacking safety avoided discussing failures, missing learning opportunities.
- Failure-friendly teams viewed setbacks as "springboards," using them to refine processes and drive innovation.

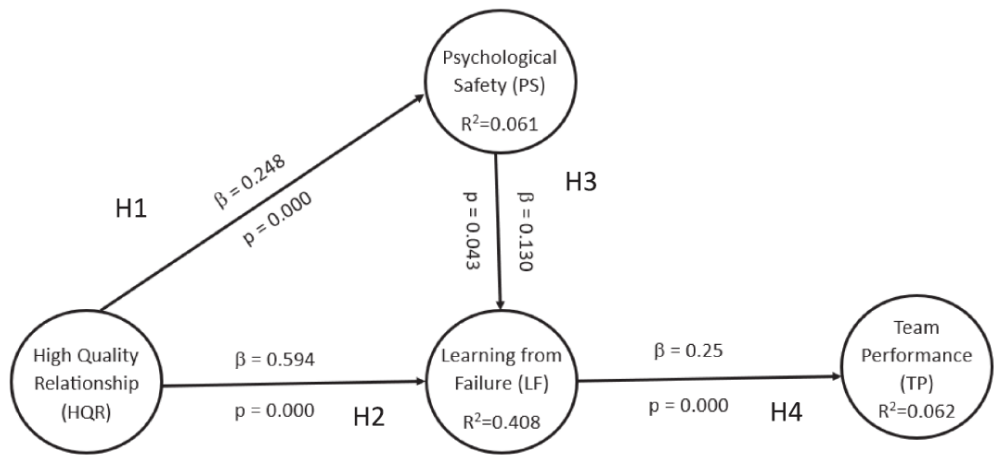

Figure 6. Studied factors and their impact on team performance.

Failure is inevitable in innovation. Teams that treat it as a learning opportunity rather than a stigma tend to adapt faster, innovate more effectively, and maintain higher morale. In large-scale ASD, where experimentation is critical, embracing failure is essential for success.

**Actionable Strategies**
- **Fail swiftly, learn deeply.** This study confirms previous research, which shows that failing is important for success. Actually, the faster you fail, the better your chances of success in any given field[2]. All winners begin as losers, and the earlier they fail, the more likely they are to become winners.
- **Empower teams with autonomy.** Involve stakeholders in deciding how to delegate decision-making to the team, which fosters ownership and accountability. For instance, allow teams to decide how to address sprint-level failures without top-down intervention.
- **Host "failure retrospectives".** Dedicate occasional retrospectives to unpacking setbacks constructively, extracting lessons for future sprints. For example, ask, "What failed during this sprint, and what can we learn from it?" in one retrospective per quarter.
- **Celebrate learning.** Recognize teams that extract valuable lessons from failures, thereby reinforcing a culture that fosters a failure-friendly environment. For instance, highlight "Lessons Learned" in company-wide newsletters.

---

2 https://www.nature.com/nature-index/news/failure-found-to-be-an-essential-prerequisite-for-success

# PEOPLE DEBT INSIGHTS

People Debt undermines skill development, morale, and retention, creating long-term vulnerabilities. Our research highlights two critical areas: skill development and employee morale.

## Skill Development – The cost of neglecting training and expertise

Insufficient training and delayed hiring impact team performance and contribute to both TD and NTD. Key findings:

- Teams lacking access to training impact individuals' ability to adopt new technologies and methodologies.
- Delayed hiring overburdens existing staff.
- High turnover disrupts knowledge continuity, as new hires often take months to reach full productivity due to insufficient onboarding support.

Skill development is crucial for staying competitive in a rapidly evolving tech landscape. Neglecting training and hiring limits an organization's ability to meet evolving demands, leading to technical debt, delays, and eventually dissatisfied customers.

### Actionable Strategies
- **Invest in continuous learning.** Allocate budget and time for training, such as workshops, certifications, or online courses. For example, offer team members five hours per sprint for self-directed learning.
- **Accelerate hiring.** Streamline recruitment processes to quickly fill skill gaps, thereby reducing the workload on existing staff.
- **Enhance onboarding.** Develop comprehensive onboarding programs, including mentorship and role-specific training, to accelerate new-hire productivity. For example, assign mentors to guide new hires through their first three sprints.
- **Measure skill gaps.** Utilize skill assessments to monitor team proficiency, pinpointing areas for targeted training.

## Employee morale – Impact of overburdened teams

Through mixed-methods research, we explored the impacts of People Debt on employee morale. Key findings:

- Overburdened developers cite burnout and frustration.
- Delayed hiring exacerbated workload issues, with teams operating at a lower capacity for months, resulting in missed deadlines and low morale.
- High turnover created a "vicious cycle," as remaining staff absorbed additional responsibilities, further eroding morale.

Low morale undermines productivity, engagement, and retention, creating a cascade of problems. Overburdened teams struggle to deliver quality work, while turnover disrupts knowledge continuity and team cohesion. In large-scale ASD, where team morale is critical for success, addressing People Debt is essential.

### Actionable Strategies

- **Balance workloads.** Monitor team capacity and redistribute tasks to prevent burnout. For example, use sprint planning to ensure no team member is assigned more than 80% capacity.
- **Prioritize hiring for critical roles.** Work with HR to expedite hiring for roles that alleviate workload pressures, such as additional developers or testers.
- **Foster work-life balance.** Encourage flexible schedules, mental health days, and recognition programs to boost morale. For instance, offer "Recharge Fridays" with no meetings.
- **Measure Morale Regularly.** Use surveys to regularly track morale, identifying early signs of burnout. Questions like "Do you feel supported in managing your workload?" can highlight issues.

# CONCLUSION

Non-Technical Debt is not a peripheral concern, it is central to the success of agile software development. The NODLA project findings make it clear: Organizational, Process, Social, and People Debts are not just inefficiencies; they are threats to team performance, innovation, and productivity. But there is hope. By implementing the strategies outlined in this white paper, organizations can manage NTD and unlock their teams' full potential.

- **Refine team structures:** Balance skills and roles to enhance knowledge sharing, cohesion, and adaptability.
- **Foster psychological safety:** Lead with transparency, normalize failure, and measure safety to boost innovation and retention.
- **Streamline processes:** Clarify roles, optimize workflows, and prioritize Process Debt to improve efficiency and satisfaction.
- **Strengthen communication:** Align teams with shared visions, build remote-friendly frameworks, and measure social health to reduce Social Debt.
- **Invest in people:** Prioritize training, accelerate hiring, and balance workloads to boost skill development and morale.

Organizations that prioritize team cohesion, psychological safety, continuous learning, and effective process management are better equipped to succeed in agile software development environments. Don't let NTD fester. By actively addressing social, process, and technical challenges, organizations can create more resilient, high-performing teams capable of adapting to the complex and dynamic nature of software development.

# FUTURE RESEARCH DIRECTIONS

While the NODLA project provides a robust foundation for understanding and addressing NTD, several areas warrant further exploration:

- **Quantitative modelling:** Develop predictive models to forecast NTD impacts on project outcomes, such as cost overruns or missed deadlines.
- **Remote and hybrid settings:** Investigate how remote work influences NTD and identify scalable solutions for distributed teams.
- **Cross-cultural analysis:** Examine how cultural factors (e.g., individualism vs. collectivism) influence NTD manifestations and mitigation strategies.
- **Longitudinal studies:** Conduct multi-year studies to track the long-term impacts of NTD interventions, such as sustained improvements in innovation or retention.
- **Integration with Technical Debt:** Explore the interplay between NTD and Technical Debt, developing holistic frameworks to address both simultaneously.
- **Advancement in technology impact:** How do generative artificial intelligence support or hinder social sustainability in LSA teams?

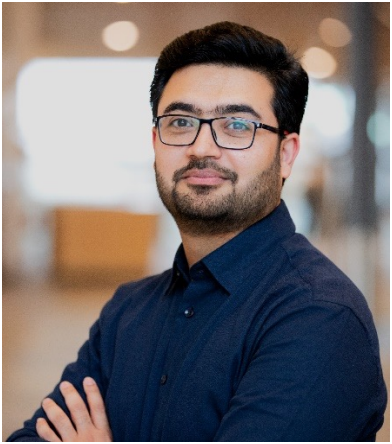

**Muhammad Ovais Ahmad**

Professor (Assoc.) in Software Engineering,
Department of Computer Science, Karlstad University, Sweden.

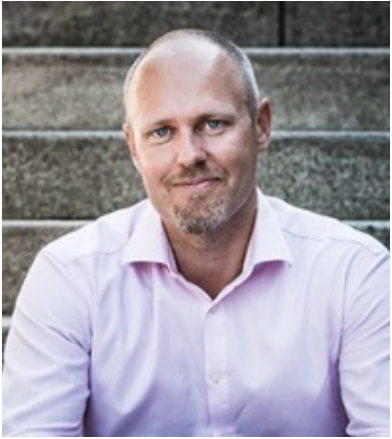

**Tomas Gustavsson**

Assistant professor of Information Systems and Project management,
Department of Informatics, Karlstad University, Sweden

KAU.SE

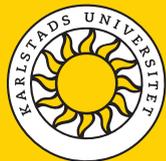